\documentclass[prl,twocolumn,showpacs,superscriptaddress,floatfix]{revtex4}
\usepackage{graphicx}
\usepackage{amsmath}
\usepackage{amssymb}
\usepackage{pifont}
\usepackage[dvips]{color}
\usepackage{subeqnarray}
\usepackage{psfrag}

\newcommand{\Db}{\mathbf{D}}

\newcommand{\Eb}{\mathbf{E}}

\newcommand{\rb}{\mathbf{r}}

\begin{document}

\title{ Double layer in ionic liquids: Overscreening vs. crowding }

\author{ Martin Z. Bazant} \affiliation{ Departments of Chemical Engineering and
  Mathematics, Massachusetts Institute of Technology, Cambridge, MA
  02139 USA}

\author{Brian D. Storey}
\affiliation{ Franklin W. Olin College of Engineering, Needham, Massachusetts 02492, USA }

\author{ Alexei A. Kornyshev }
\affiliation{ Department of Chemistry, Imperial College London, SW7 2AZ London, U.K. }

\date{\today}

\begin{abstract}

We develop a simple Landau-Ginzburg-type continuum theory of solvent-free ionic liquids and use it to predict the structure of the electrical double layer. The model captures overscreening from short-range correlations, dominant at small voltages, and steric constraints of finite ion sizes, which prevail at large voltages.  Increasing the voltage gradually suppresses overscreening in favor of the crowding of counterions in a condensed inner layer near the electrode. The predicted ion profiles and capacitance-voltage relations are consistent with recent computer simulations and experiments on room-temperature ionic liquids, using a correlation length of order the ion size.

      \end{abstract}

      \pacs {}

\maketitle

{\it Introduction. ---}
The rediscovery of room temperature ionic liquids (RTILs)  as designer solvents promised a revolution in synthetic chemistry \cite{welton1999}.  Thousands of RTILs have  been synthesized  with large organic cations and similar organic or smaller inorganic anions. Non-volatile and capable of withstanding up to $\pm 4\mbox{-}6$ V without decomposition, RTILs also hold  promise as solvent-free electrolytes for super-capacitors, solar cells, batteries and electroactuators \cite{silvester2006,simon2008,freyland2008,armand2009,ito2008,bai2008,buzzeo2004,ye2001,bhushan2008}.

\begin{figure}
\begin{center}
\includegraphics[width=2.7in]{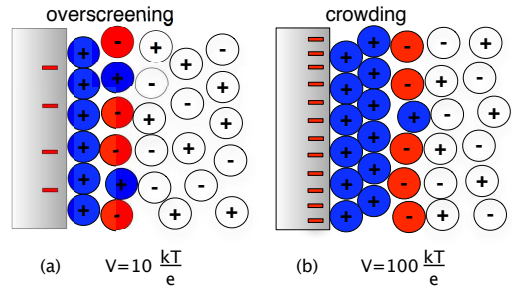}
\caption{ Structure of the ionic-liquid double layer (in color) predicted by our theory and molecular dynamics simulations~\cite{fedorov2008} (Figs. 2-3 below). (a) At a moderate voltage, $V = 10 k_BT/e$ (0.26 V), the surface charge is overscreened by a monolayer of counterions, which is corrected by an excess of co-ions in the second monolayer. (b) At a high voltage, $V = 100 k_BT/e$ (2.6 V), the crowding of counter-ions extends across two monolayers and dominates overscreening, which now leads to a co-ion excess in the third monolayer. Due to electrostriction, the diffuse double layer (colored ions) is more dense than the quasi-neutral bulk liquid (white ions).
\label{fig:cartoon} }
\end{center}
\end{figure}

For such applications, it is crucial to understand the structure of the RTIL/electrode double layer. The classical Gouy-Chapman-Stern (GCS) model for dilute electrolytes was used to interpret RTIL capacitance data until recently, when a mean-field theory for the {\it crowding} of finite-sized ions~\cite{kornyshev2007} suggested bell or camel shapes of the differential capacitance versus voltage, decaying as  $C \sim V^{-1/2}$. These were basically confirmed in subsequent experimental~\cite{alam2007,alam2007b,alam2008,lockett2008,zhou2010}, theoretical~\cite{oldham2008,lauw2009}  and computational\cite{fedorov2008,fedorov2008b,fedorov2010,trulsoon2010,vatamanu2010} studies. Similar theories have also been developed for highly concentrated electrolytic solutions~\cite{large_acis,kilic2007a,freise1952}, but none of these models accounts for short-range Coulomb correlations~\cite{levin2002}, which could be very strong in RTIL~\cite{mezger2008,skinner2010}. As first revealed by linear response theories of molten salts \cite{rovere1986}, correlations generally lead to {\it over-screening}~\cite{levin2002}, where the first layer at the electrode delivers more counter-charge than is on the surface; the next layer then sees a  smaller net charge of the opposite sign, which it again overscreens; and so-on, until neutrality is reached. Recent computer simulations of a model RTIL/electrode interface have demonstrated overscreening structures at low voltage, similar to experiments~\cite{mezger2008}, which are gradually overcome by the formation of a condensed layer of counter-ions at high voltage ~\cite{fedorov2008}, as shown in Fig. ~\ref{fig:cartoon}.

In this Letter, we suggest a phenomenological theory to describe the interplay between over-screening and crowding. Compared to more involved models of statistical mechanics, the theory only crudely approximates discrete interactions near a surface, but it is simple enough to be applied to dynamical problems in nanotribology, electroactuation, and porous super-capacitors.

{\it Theory. --- } We propose a Landau-Ginzburg-like functional for the total free energy~\cite{EPAPS}:
\begin{equation}
G = \int_V d\rb \left\{ g + \rho\phi - \frac{\varepsilon}{2} \left[ |\nabla\phi|^2  + \ell_c^2 (\nabla^2\phi)^2\right] \right\}
+ \oint_S d\rb \, q_s \phi  \label{eq:G}
\end{equation}
where $g(c_+,c_-)$ is the enthalpy density, depending on the ionic concentations $c_\pm$, as described below; $\rho = e(z_+ c_+ - z_- c_-)$ is the mean charge density in the liquid volume $V$; $q_s$ is the surface charge density on a bounding metal surface $S$; $\phi$ is the mean electrostatic potential, and we subtract the self energy of the electric field $- \frac{\varepsilon}{2} |\nabla\phi|^2$, assuming a constant permittivity $\varepsilon$ to describe the polarizability of the ions. The first three terms in brackets  are those  used in  mean-field theories of ionic liquids~\cite{kornyshev2007}, ionic crystals~\cite{grimley1947} and electrolytes~\cite{large_acis,borukhov1997}. To go beyond that approximation, we introduce the next allowable potential-gradient term, $- \frac{\varepsilon}{2} \ell_c^2 (\nabla^2\phi)^2$, similar to Cahn-Hilliard concentration-gradient expansions ~\cite{cahn1958,nauman2001}, where $\ell_c$ is an electrostatic correlation length~\cite{EPAPS}.

The sign of the correlation term is negative to describe {\it over-screening} in strongly correlated liquids: The energy is lowered by {\it enhancing} the curvature of $\phi$, a measure of the ``mean-field charge density", $\bar{\rho}\equiv -\varepsilon\nabla^2\phi$. For point charges,  $\ell_c$ is on the order of the Bjerrum length $\ell_B=(ze)^2/4\pi\varepsilon kT$ (in SI units). For RTILs with $\varepsilon\approx 10 \varepsilon_0$, the Bjerrum length, $\ell_B\approx 5.5$ nm,  is much larger than the ion diameter, $a\approx 1$ nm ~\cite{welton1999}, so the correlation length $\ell_c \approx a$ is typically at the molecular scale ~\cite{mezger2008}.

Setting $\delta G/\delta \phi = 0$ for bulk and surface variations ~\cite{EPAPS}, 
we obtain a modified Poisson equation
\footnote{ Similar equations have been derived for the equilibrium profile of point-like counterions near a charged wall; C. Santangelo [Phys. Rev. E {\bf 73}, 041512 (2006)] showed that (\ref{eq:Poisson}) is exact for both weak and strong  coupling and  a good approximation at intermediate coupling with $\ell_c = \ell_B$; M. M. Hatlo and L. Lue [Europhys. Lett. {\bf 89}, 25002 (2010)] developed a systematic approximation for $\ell_c$.  }
\footnote{ A. Hildebrandt, R. Blossey, S. Rjasanow, O. Kohlbacher, and H.-P. Lenhof [Phys. Rev. Lett. {\bf 93}, 108104 (2004)] derived a similar gradient approximation for nonlocal solvent dielectric response [A. Kornyshev, A. I. Rubinstein, and M. A. Vorotyntsev, J. Phys. C {\bf 11}, 3307 (1978)].}
and modified electrostatic boundary condition, respectively:
\begin{eqnarray}
 \varepsilon (\ell_c^2 \nabla^2 - 1)  \nabla^2 \phi &=& \rho = \nabla \cdot \Db,    \label{eq:Poisson} \\
 \hat{n}\cdot\varepsilon(\ell_c^2 \nabla^2 - 1)\nabla \phi &=& q_s = \hat{n}\cdot\Db,
  \label{eq:PBC}
\end{eqnarray}
where $\Db$ is the displacement field.  Due to correlations, the medium permittivity $\hat{\varepsilon}$, defined by $\Db = - \hat{\varepsilon}\nabla\phi$, is a linear differential operator,
%\begin{equation}
$\hat{\varepsilon} = \varepsilon \left( 1 - \ell_c^2 \nabla^2 \right)$,
%  \label{eq:eps}
%\end{equation}
whose Fourier transform (valid for wavenumber $|k| \ll \ell_c^{-1}$), $\hat{\varepsilon}_k \sim \varepsilon( 1 + \ell_c^2 k^2)$, increases with $k$, as is typical for molten salts~\cite{tosi1991}.
It is important to note that our $\hat{\varepsilon}$  is not the complete dielectric function of the ionic liquid, which should diverge at small
$k$, as for any conducting medium \cite{tosi1991}.
This divergence is subtracted since translational degrees of freedom are treated explicitly via $\rho(\phi)$, which also  takes into account the nonlinear response in the rearrangement of ions. In our model, $\hat{\varepsilon}$ approximates the linear dielectric response of the liquid of correlated ion pairs (zwitterions), which are  considered to be bound by stronger forces, independent of the mean electric field.

Since Poisson's equation (\ref{eq:Poisson}) is now  fourth-order, we need additional boundary conditions,
similar to  electrodynamics with spatial dispersion~\cite{agranovich1984}. Consistent with our bulk gradient expansion,  we neglect correlations at the surface
and apply the standard boundary condition, $-\varepsilon\hat{n}\cdot\nabla\phi = q_s$.  Equation (\ref{eq:PBC}) then implies
%\begin{equation}
$\hat{n}\cdot \nabla (\nabla^2 \phi) = 0$,  %\label{eq:extraBC}
%\end{equation}
which requires that the mean-field charge density is ``flat" at the surface, $\hat{n}\cdot \nabla \bar{\rho} = 0$, consistent with a continuum model of finite-sized ions.

Following Ref.~\cite{kornyshev2007}, we  describe crowding effects via the classical model~\footnote{This model was originally developed for concentrated electrolytes~\cite{bikerman1942} and ionic solids~\cite{grimley1947}, as reviewed in \cite{large_acis}.}:
\begin{eqnarray}
g &=& \frac{k_B T}{v} \left\{ v c_+ \ln (v c_+) + v c_- \ln (v c_-) \right. \nonumber \\
& + &  \left. \left[1 - v(c_+ + c_-)\right] \ln \left[1 - v(c_+ + c_-)\right] \right\} \label{eq:lattice}
\end{eqnarray}
which is the entropy density $g=-TS/v$ of an ideal solution of cations, anions, and holes, respectively, of minimum volume $v$. We set $v=(\pi/6)a^3/\Phi_{max}=0.83 a^3$ for random close packing of spheres at volume fraction $\Phi_{max}=0.63$.  More accurate expressions for $g$ are available for uniform hard-sphere mixtures~\cite{large_acis}, but, due to the breakdown of the local-density approximation~\cite{levin2002}, they over-estimate steric repulsion in the double layer~\cite{antypov2005}. The  weaker  repulsion in (\ref{eq:lattice}) actually  provides a better first approximation for the packing entropy.

\begin{figure}
\begin{center}
\includegraphics[width=2.75in]{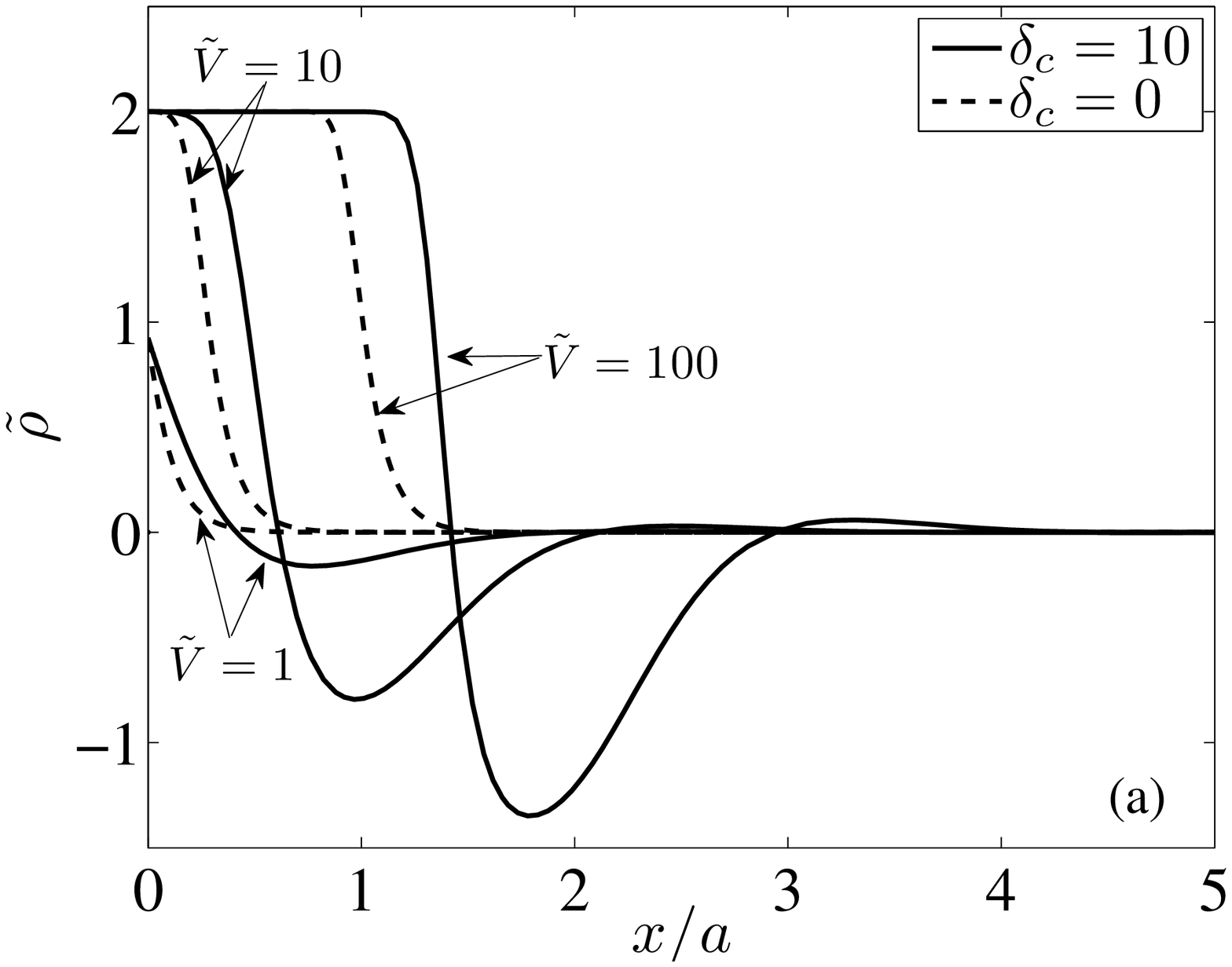}\\
\includegraphics[width=2.75in]{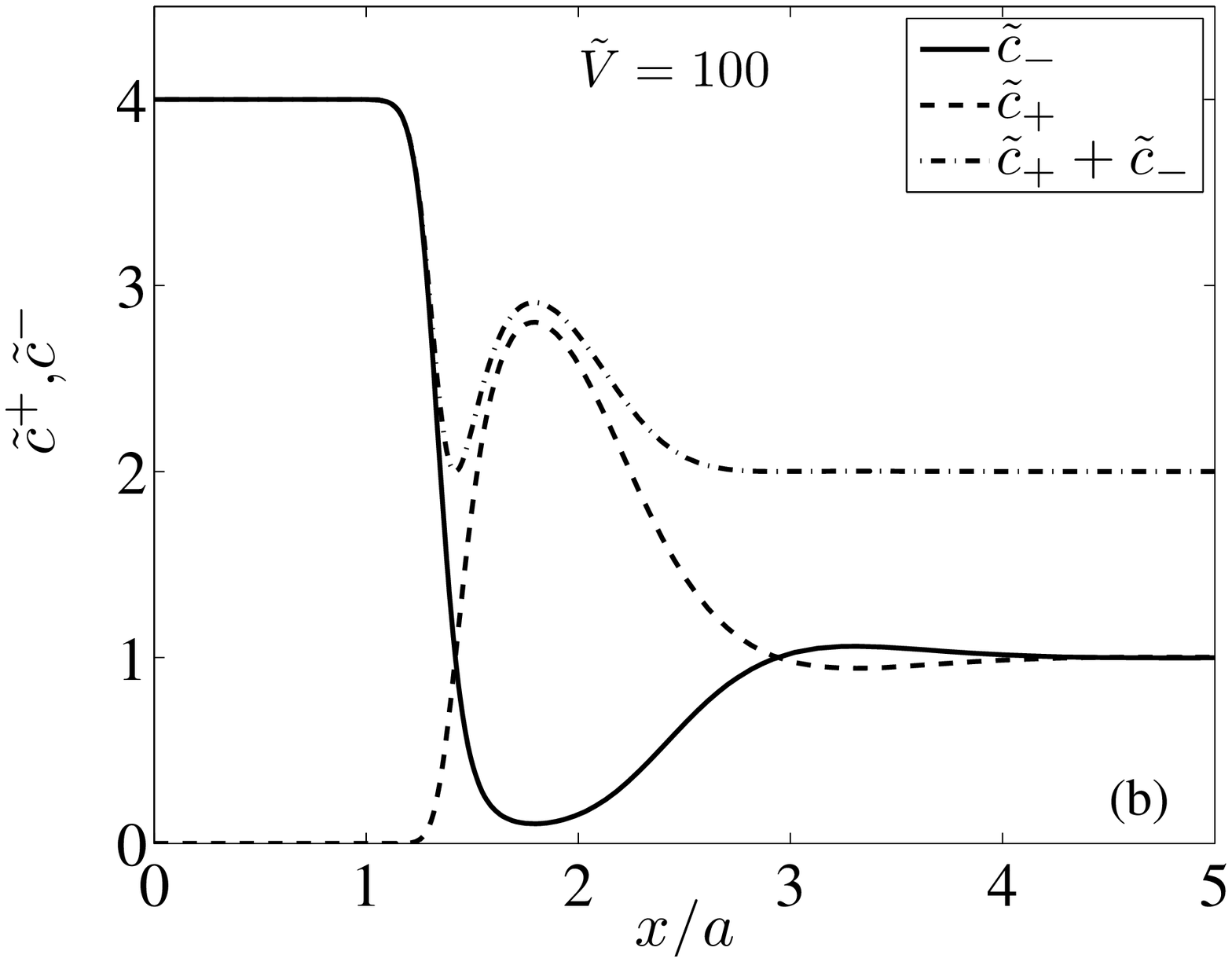}
\caption{
Voltage-dependent  double-layer structure predicted by our model.
 (a) Dimensionless charge density versus $x/a$ for $\tilde{V}=eV/k_BT=1, 10 ,100$  (solid curves), compared to the mean-field theory~\cite{kornyshev2007} with $\delta_c=0$  (dashed curves).
(b) Dimensionless cation (solid) and anion (dashed) concentrations and mass density (dash-dot) at high voltage, $\tilde{V}=100$.
Position $x$ is measured from the distance of closest approach and scaled to the ion diameter $a=10$\AA. Model parameters $\gamma=0.5$ (bulk/maximum density), $\delta_c=10$ (correlation/Debye length), and  $\varepsilon=5 \varepsilon_0$ are estimated from ion profiles in simulations ~\cite{fedorov2008} (Fig. 3 below).
\label{fig:rho}  }
\end{center}
\end{figure}

The electrochemical potentials of the ions are then
\begin{equation}
\mu_\pm  = \frac{ \delta G}{\delta c_\pm} = k_B T \ln\left[ \frac{ c_\pm }{1 - v(c_+ + c_-)}\right]   \pm z_\pm e \phi,
\end{equation}
and their gradients $\nabla\mu_\pm$ produce ionic fluxes~\cite{large_acis}. In equilibrium with a reference solution with $\phi=0$ and volume fraction,
%\begin{equation}
$\gamma =  2vc_+^{ref}=2vc_-^{ref}$,
%\end{equation}
the conditions $\mu_\pm=$constant determine the Fermi-like charge density distribution, $\rho(\phi)$. In electrolytes, $\gamma$ is the volume fraction of solvated ions in the bulk
~\cite{bikerman1942,borukhov1997,kilic2007a,large_acis}. In ionic liquids $\gamma$ ($\leq 1$) is the ratio of the bulk ion density to the maximum possible density,  which characterizes their ability to compress~\cite{kornyshev2007}. In equilibrium, we obtain a (dimensionless) modified Poisson-Fermi equation,
%\begin{equation}
%\varepsilon (1 - \ell_c^2 \nabla^2)\nabla^2 \phi = \frac{ze}{v}
%\frac{ \sinh(ze\phi / k_B T) }
%{1+2\gamma\sinh^2(ze\phi / 2 k_B T) }
%\end{equation}
\begin{equation}
(1 - \delta_c^2 \tilde{\nabla}^2)\tilde{\nabla}^2\tilde{\phi} = \frac{ \sinh \tilde{\phi}}{1+2\gamma\sinh^2(\tilde{\phi}/2) } = -\tilde\rho(\tilde{\phi})
\end{equation}
where $\tilde{x} = x/\lambda_D$, $\tilde{\nabla}=\lambda_D\nabla$, $\tilde{\phi} = ze\phi/k_BT$. Here, $\lambda_D = \sqrt{\varepsilon k_BT v}/ze$ is the Debye screening length, and $\delta_c = \ell_c/\lambda_D$ is the dimensionless correlation length, which controls deviations from the mean-field theory.
%Note that $-\tilde{\rho}=\tanh\tilde{\phi}$ for $\gamma=1$ ~\cite{kornyshev2007,oldham2008}.
For $\varepsilon=10 \varepsilon_0$ and $a = 10$ \AA, the Debye length is very small, $\lambda_D = 1.1$ \AA, so the ion size $a$ becomes the relevant length scale~\footnote{ Due to the local-density approximation, our model cannot resolve discrete layers of ions~\cite{large_acis}, but more accurate  weighted-density approximations~\cite{levin2002} require solving nonlinear integro-differential equations. }. If we chose $\delta_c = 10$ to reproduce double-layer properties from simulations~\cite{fedorov2008} (below), then correlations are indeed at the molecular scale, $\ell_c \approx a$.

{\it Results. ---} Let us apply our model to a half space by solving $\delta_c^2 \tilde{\phi}^{\prime\prime\prime\prime} - \tilde{\phi}^{\prime\prime} =  \tilde{\rho}(\tilde{\phi})$ for $\tilde{x}>0$ subject to $\tilde{\phi}^{\prime\prime\prime}(0)=0$, and $\tilde{\phi}(0)=\tilde{V}=zeV/k_BT$, where $V$ is the surface potential relative to the bulk. We solve the model analytically for small, moderate and large voltages \cite{EPAPS} and compare with numerical solutions.

\begin{figure}
\begin{center}
\includegraphics[width=\linewidth]{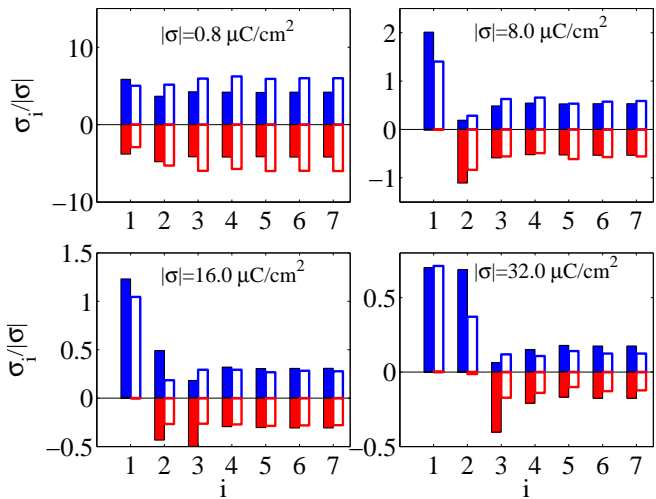}
\caption{ Distributions of cations (above) and anions (below) sorted into monolayer bins $i=1,2,\ldots$ for different surface charges $\sigma$, as predicted by our model  (solid bars) in qualitative agreement with simulations (Fig. 2 of Ref.~\cite{fedorov2008}, open bars).
\label{fig:layers}  }
\end{center}
\end{figure}

\underline{ 1. Structure of the double layer.}
In Fig. \ref{fig:rho} we show the calculated charge density (a), mass density and ion concentrations (b) for $\gamma=1/2$ and $\delta_c=10$. For $a=10$ \AA, $T =450 K$ and $\varepsilon=5\varepsilon_0$, which imply $\ell_c = 0.95 a$, the model predicts molecular-scale charge-density oscillations, similar to experiments~\cite{mezger2008} and in good agreement with simulations~\cite{fedorov2008}, as shown in Fig.~\ref{fig:layers}. At small potentials, the oscillation period and damping length are $\tilde{\lambda}_o\sim2\pi\sqrt{2\delta_c}$ for $\delta_c\gg 1$ \protect\cite{EPAPS}, or with units restored, $\lambda_o\sim 2\pi\sqrt{\lambda_D \ell_c} =  20$ \AA$= 2.0 a$.  With increasing voltage, a condensed layer of counterions forms and expands into the bulk, as predicted by the mean-field theory~\cite{kornyshev2007}, but with the important difference that this layer overscreens the surface charge, leading to a second layer of excess co-ions, which again (slightly) overscreens and triggers the same low-voltage damped charge-density oscillations.
%This spatial variation of overscreening can be confirmed by integrating the charge density~\cite{EPAPS}.
%At very large voltages, the crowding of counterions in the first layer dominates overscreening in %determining double-layer properties.
The model also predicts non-uniform electrostriction at high voltage (Fig. \ref{fig:rho}(b)) consistent with simulations (Fig.~\ref{fig:layers}): The first counterion layer attains the maximum density, while the next co-ion-rich layer has a lower density, but still larger than the bulk.

\begin{figure}
\begin{center}
\includegraphics[width=3.0in]{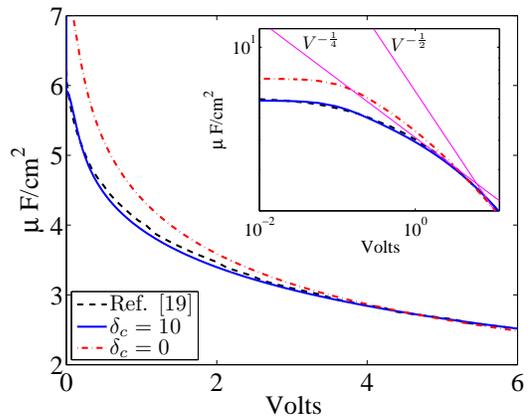}
\caption{ Double-layer differential capacitance $C_{d}$ from our model (solid), simulations~\cite{fedorov2008} (dashed),  mean-field theory~\cite{kornyshev2007} (dash-dot), and  our asymptotic scalings (inset).
\label{fig:cap} }
\end{center}
\end{figure}

\underline{ 2.  Double-layer capacitance. }
An important property of the double layer is its voltage-dependent capacitance $C(V)$. It has been found that excluded volume effects  explain trends in the experimental data, but the mean-field theory over-estimates $C$,  unless an empirical Stern-layer correction is added~\cite{fedorov2008,fedorov2008b}. In Fig. \ref{fig:cap} we show the double-layer capacitance versus voltage in our model, which is in very close agreement with simulations of Ref. ~\cite{fedorov2008} without fitting any additional parameters. We only account for the extra capacitance, $C_s = 2 \varepsilon/a$, in series with the diffuse double layer, due to the distance of closest approach of ion centers, $a/2$.
The value of $C_s$ relative to the mean-field Debye value, $C_D=\varepsilon/\lambda_D$,
is $\tilde{C}_s = C_s /C_D = 2 \lambda_D/a \approx 2/\delta_c$.

At low-voltage, the model can be linearized and solved to find
the diffuse layer capacitance, $C_{d}$ \cite{EPAPS},
%correlations reduce  the diffuse layer $C_{DL}$
%compared to $C_D$ by 60\% for $\delta_c=10$.
%More generally, by solving the linearized model, we find [42]:
\begin{equation}
\tilde{C}_{d} = \frac{ C_{d} \lambda_D}{\varepsilon}
\sim  \frac{\sqrt{2 \delta_c + 1}}{ \delta_c+1} \ \ \mbox{ for } |\tilde{V}| \ll 1.
%~~ \mathrm{At~low~voltage}.
\label{eq:Cap_lowV}
\end{equation}
%where $C$ decreases as the square root of the correlation length.
% This formula provides a simple way to extract the correlation length from capacitance data. For example,
% simulations with $\tilde{C}(0)=0.15$  (Fig. 4 of \cite{fedorov2008}) imply $\delta_c = 88$.
By extending  the Composite Diffuse Layer Model of Ref. ~\cite{kilic2007a} we can also approximate  $C_{d}$  at
moderate voltages, once the condensed counterion layer forms
and $\delta_c \gg 1$;
\begin{equation}
\tilde{C}_{d}  \sim \frac{ 8^{3/4} }{ 3  (\delta_c^2 \gamma \tilde{V})^{\frac{1}{4}} }
 \ \ \mbox{ for } \ \  \frac{128}{81 \gamma} \ll  |\tilde{V}| \ll  \frac{81}{128 \gamma \delta_c^2}.
% ~~ \mathrm{When~a~condensed~layer~forms}.
\label{eq:Cap_condensed}
\end{equation}
This  scaling breaks down at very large voltages
when the condensed layer of charge grows enough to dominate the capacitance, yielding $\tilde{C}_{d} \sim \sqrt{{2}/{\gamma \tilde{V}}}$
as in the mean-field theory
\cite{kornyshev2007,kilic2007a}.
These scalings compare well with numerical solutions for $\delta_c \gg 1$ \cite{EPAPS} and explain why our model is closer to simulations than the mean-field theory without correlations (Fig. \ref{fig:cap}).
%%, Eq. \ref{eq:Cap_lowV} and \ref{eq:Cap_condensed} (dashed lines).
%The crossover between the regimes is given by the intersection of the three capacitance relationships as
%seen clearly when $\delta_c=10$ in Fig. \ref{fig:rho}c.
%Formally,  the curves for all
%$\delta_c$ approach the $\delta_c=0$ limit at high voltage,
%but this regime is of little practical interest when $V \gg 6$ Volts.

{\it Conclusion. ---}
In this paper we have made a first attempt to describe both overscreening and crowding in dense Coulomb liquids, such as RTILs and molten salts.  Our simple phenomenological theory predicts that overscreening is  pronounced at small voltages and gradually replaced by the formation of a condensed layer of counterions, followed by complete lattice saturation at very large voltages.  Each of these three regimes is characterized by its own capacitance-voltage dependence. Our findings are in line with simulations and experiments, and they give a more complete picture of the nonlinear polarization of ionic liquids.

\section*{Acknowledgments}
This work was supported by the National Science Foundation, under contracts DMS-0707641 (MZB) and CBET-0930484 (BDS), and the Engineering and Physical Sciences Resarch Council under grant EP/H004319 (AAK). It was influenced by joint work with M. Fedorov (AAK) and discussions with A. Maggs.

\bibliography{elec28c}

\clearpage

\begin{center}
{\large {\bf SUPPLEMENTARY INFORMATION }}
\end{center}

\vspace{0.2in}

\section{ Phenomenological theory of electrostatic correlations }

Let $G = G_{el} + G_{chem}$, where $G_{el}$ is the electrostatic energy and $G_{chem} = \int_V d\rb g$ is the chemical (non-electrostatic) part of the free energy.  Suppose that $G_{chem}$ is known, and focus on electrostatic correlation effects in $G_{el}$.

The electrostatic potential,  $\phi$, is the free energy per ion (free charge).  The  electrostatic energy cost for adding a charge $\delta \rho$ in the bulk liquid volume $V$ or $\delta q_s$ on the metal surface $S$ is,
\begin{equation}
\delta G_{el} = \int_V d\rb \, \phi \, \delta \rho + \int_S d\rb \, \phi \, \delta q_s.   \label{eq:dG}
\end{equation}
The charge is related to the displacement field $\Db$ via Maxwell's ``first" equation,
\begin{equation}
\nabla \cdot \Db = \rho \ \ \Rightarrow \ \ \delta \rho = \nabla \cdot \delta \Db.    \label{eq:max1}
\end{equation}
The corresponding boundary condition for an ideal metal surface (where $\Db=0$) is,
\begin{equation}
[\hat{n} \cdot \Db] = \hat{n}\cdot \Db = - q_s \ \ \Rightarrow \ \ \delta q_s = - \hat{n} \cdot \Db.   \label{eq:max2}
\end{equation}
Substituting these expressions into (\ref{eq:dG}) and using Gauss' theorem, along with the definition of the electric field, $\Eb = - \nabla\phi$, we recover the standard electrostatic free energy equation~\cite{landau},
\begin{equation}
\delta G_{el} = \int_V d\rb \, \Eb \cdot \delta \Db.  \label{eq:dG2}
\end{equation}

In the linear response regime (for small external electric fields), we have
\begin{equation}
\Db = \hat{\varepsilon} \Eb,
\end{equation}
where $\hat{\varepsilon}$ is a linear operator, whose Fourier transform $\hat{\varepsilon}(k)$ encodes how  {\it the permittivity depends on the wavelength} $2\pi/k$ of the $k$-Fourier component of the field, due to discrete ion-ion correlations, as well as any non-local dielectric response of the ions, such as exponentially decaying Debye correlations in ionic plasma, as well as correlations in polarization flucutations due to any other molecules if they are present in the liquid. We can then integrate  (\ref{eq:dG2}) over $\delta \Db$ through a charging process that creates all the charges in the bulk and surface from zero to obtain
\begin{equation}
G_{el} = \frac{1}{2} \int_V d\rb \, \Eb \cdot \Db.
\end{equation}

For a given distribution of charges $\rho$ and $q_s$, with associated displacement field $\Db$, the physical electric field $\Eb$ is the one that minimizes $G_{el}$, subject to the constraint of satisfying Maxwell's equations (\ref{eq:max1})-(\ref{eq:max2}). Since $\Eb = -\nabla\phi$ to enforce $\nabla\times\Eb=0$, we can minimize $G_{el}$ with respect to variations in $\phi$, using Lagrange multipliers $\lambda_1$ and $\lambda_2$ to enforce the constraints,
\begin{eqnarray}
G_{el}[\phi] &=& \int_V  d\rb \, \left[ \frac{1}{2} \Eb\cdot\Db+ \lambda_1 \left(\rho - \nabla\cdot \Db\right)\right] \nonumber \\
& & + \oint_S d\rb_s \, \lambda_2 \left( q_s + \hat{n}\cdot\Db \right).
\end{eqnarray}
To calculate the extremum, we use the
Fr\'echet functional derivative:
\begin{equation}
\frac{\delta G_{el}}{\delta \phi} = \lim_{\epsilon\to 0} \frac{ G_{el}[\phi + \epsilon \phi_0 \delta_\epsilon] - G_{el}[\phi] }{\epsilon \phi_0}
\end{equation}
where $\delta\phi_\epsilon = \phi_o \delta_\epsilon(\rb,\rb^\prime)$ is a localized perturbation of the potential (with compact support), which tends either to a 3D delta function in the liquid ($\rb \in V$) or to a 2D delta function on the surface ($\rb \in S$) as $\epsilon\to 0$, and $\phi_0$ is an arbitrary potential scale for dimensional consistency. By setting  ${\delta G_{el}}/{\delta \phi} = 0$ for both surface and bulk variations, we find $\lambda_1=\lambda_2=\phi$. Finally, using vector identities, we arrive at a general functional for the electrostatic energy,
\begin{equation}
G_{el}[\phi] = \int_V  d\rb \, \left( \rho\phi +  \frac{1}{2} \nabla\phi \cdot\Db \right) + \oint_S d\rb_s \, q_s \phi
\end{equation}
to be minimized with respect to $\phi$, once we know the relationship between $\Db$ and $\Eb = -\nabla\phi$.

To model the field energy in an ionic liquid, we assume {\it linear dielectric response of the molecules with constant permittivity $\varepsilon$ plus a non-local contribution for ion-ion correlations}.
Here, the permittivity $\varepsilon$ describes the electronic polarizability of the ions.
\begin{equation}
g_{field} = -\frac{1}{2} \nabla\phi \cdot\Db =  \frac{\varepsilon}{2}  \left( \Eb(\rb)^2 + \int_V d\rb^\prime K(\rb,\rb^\prime) \bar{\rho}(\rb)\bar{\rho}(\rb^\prime) \right)
\end{equation}
where
\begin{equation}
\bar{\rho} = \varepsilon \nabla\cdot \Eb = -\varepsilon \nabla^2\phi,
\end{equation}
is the ``mean-field charge density'', which would produce the electric field in the dielectric medium without accounting for ion-ion correlations.
Suppose that the non-local kernel $K(\rb,\rb^\prime)$ decays over a length scale $\ell_c$, bounded below by the finite ion size $a$ and above by the Bjerrum length $\ell_B$, which sets the scale for electrostatic correlations among point charges. For charge variations over scales larger than $\ell_c$ (corresponding to small perturbation wavenumbers, $\ell_c |k| \ll 1$), we obtain a  gradient expansion for the non-local term
\begin{equation}
g_{field} \sim \frac{\varepsilon}{2} \left[ |\nabla \phi |^2 +
\sum_{n=0}^\infty \alpha_{n}  \left( \frac{\ell_c^{n-1}}{\varepsilon} \nabla^n \bar{\rho}\right)^2  \right]
\end{equation}
where $\alpha_n$ are dimensionless coefficients, which implies
\begin{eqnarray}
G_{el}[\phi] &\sim& \int_V  d\rb \, \left\{ \rho\phi - \frac{\varepsilon}{2} \left[  |\nabla\phi|^2
+  \sum_{n=2}^\infty \alpha_{n-2} (\ell_c^{n-1} \nabla^{n} \phi)^2 \right] \right\} \nonumber \\
& & + \oint_S d\rb_s \, q_s \phi    \label{eq:Ggen}
\end{eqnarray}
Equation (1) in the main text results from the first term in the gradient expansion of the non-local electrostatic energy with the choice $\alpha_0=1$ (after suitably rescaling $\ell_c$), where the overall negative sign of this term is chosen to promote over-screening.

By settting ${\delta G_{el}}/{\delta \phi} = 0$ for bulk and surface perturbations in (\ref{eq:Ggen}), we recover Maxwell's equations (\ref{eq:max1})-(\ref{eq:max2}), with $\Db = \hat{\varepsilon} \Eb$, where the permittivity operator has the following gradient expansion,
\begin{equation}
\hat{\varepsilon} = \varepsilon \left( 1-  \sum_{n=1}^\infty \alpha_{n-1} \ell_c^{2n} \nabla^{2n} \right)
\end{equation}
and corresponding small-$k$ expansion of the Fourier transform,
\begin{eqnarray}
\hat{\varepsilon}(k) &=& \varepsilon \left[ 1 + \sum_{n=1}^\infty \alpha_{n-1} (-1)^{n-1} (\ell_c k)^{2n} \right] \\
& \sim & \varepsilon \left[ 1 + \alpha_0 (\ell_c k)^2 \right]
\end{eqnarray}
which grows with $k$ at small wavenumbers in the case where correlations promote overscreening, $\alpha_0>0$.
Note that it is well known that such an expansion only holds at small $k$. At larger $k$, $\hat{\varepsilon}(k)$ diverges, becomes negative on the other side of the singularity, then diverges again to $-\infty$ at another point, and becomes positive after the second divergence; see Refs. \cite{tosi1991,rovere1986}.

\section{ Charge profiles and verification of over-screening  }

In the main text we show charge density profiles for a specific set of parameters.
 The charge density profiles in the text are presented in spatial coordinates scaled by the ion size.
However,  the natural length scale for the dimensionless problem is the Debye length.
 The solutions to the equation in dimensionless form  depend upon the applied voltage, the correlation length scale $\delta_c$, and the volume
fraction $\gamma$. In Fig. \ref{fig:rho} we show the charge density as a function of distance (normalized by the Debye length) for increasing values of $\delta_c$. To convert these ion profiles to dimensional form, the $x$-axis
need only be scaled by the value of $\lambda_D/a$ as given by
the physical parameters of the problem.
Fig. \ref{fig:rho} shows that the strength of the over-screening is a strong function of $\delta_c$.

To prove that our simple continuum model predicts over-screening by the first, condensed layer of counterions, in Fig. \ref{fig:rho_integrated} we plot the integrated charge density up to position $x$ from the surface versus $x$. The integrated charge is then normalized by the
total charge in the double layer, as in Ref. ~\cite{fedorov2008}.
This graph  provides a quantitative characterization of the strength of over-screening in the first layer.

\begin{figure*}
\includegraphics[width=2.3in]{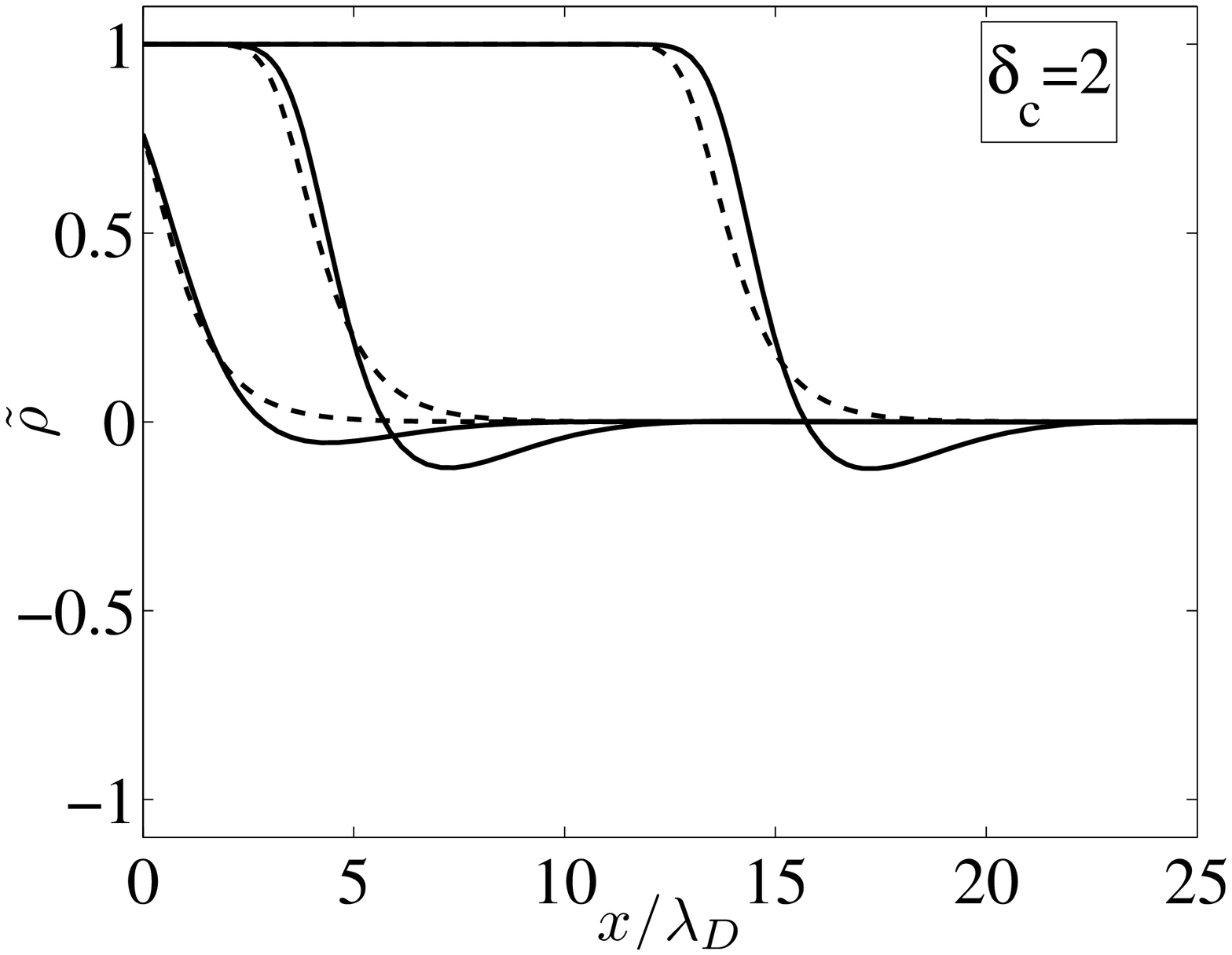}
\includegraphics[width=2.3in]{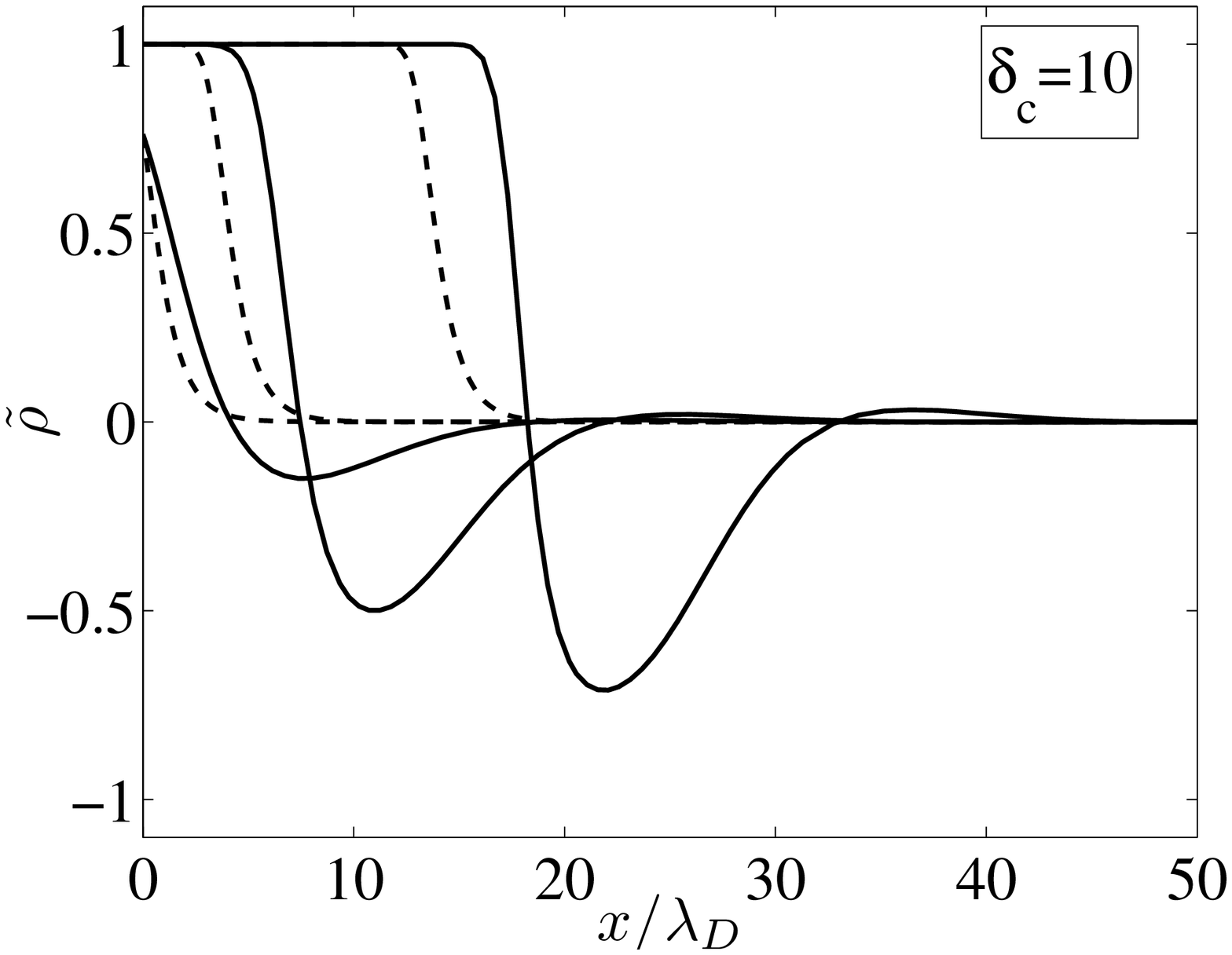}
\includegraphics[width=2.3in]{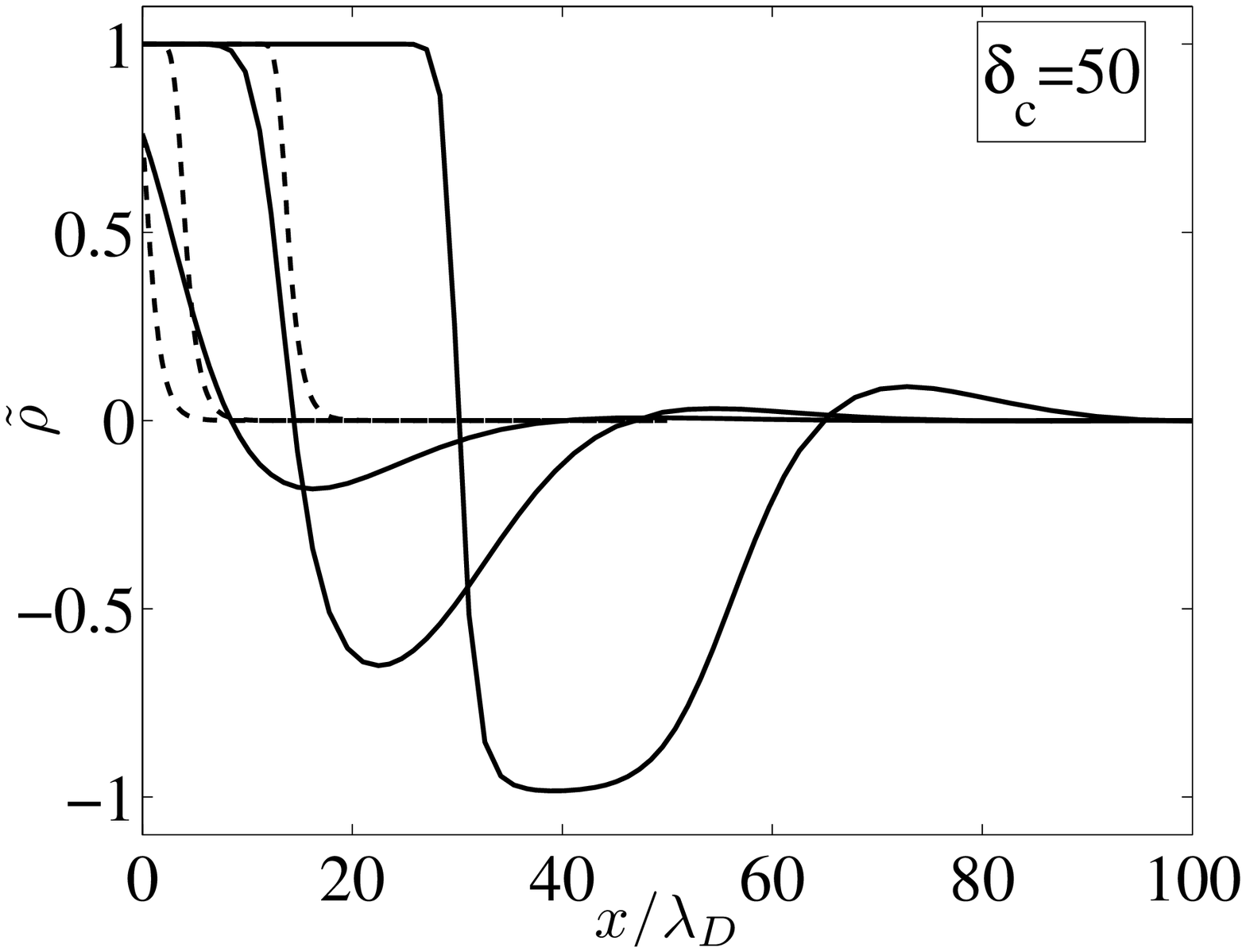}
\caption{Charge density profiles for $\gamma=1$ and
$\delta_c=2, 10,~\mathrm{and}~50$, as indicated. The solid curves are calculated from our model while the
dashed curves are calculated for the case where  $\delta_c=0$ and there are no correlations.
Solutions are shown  for applied voltages of $\tilde{V}=1,10$ and $100$ measured in units of $k_B T/e \approx 25$ mV. Similar results are obtained for different values of the volume fraction, $\gamma$. }
\label{fig:rho}
\end{figure*}

\begin{figure*}
 \includegraphics[width=2.3in]{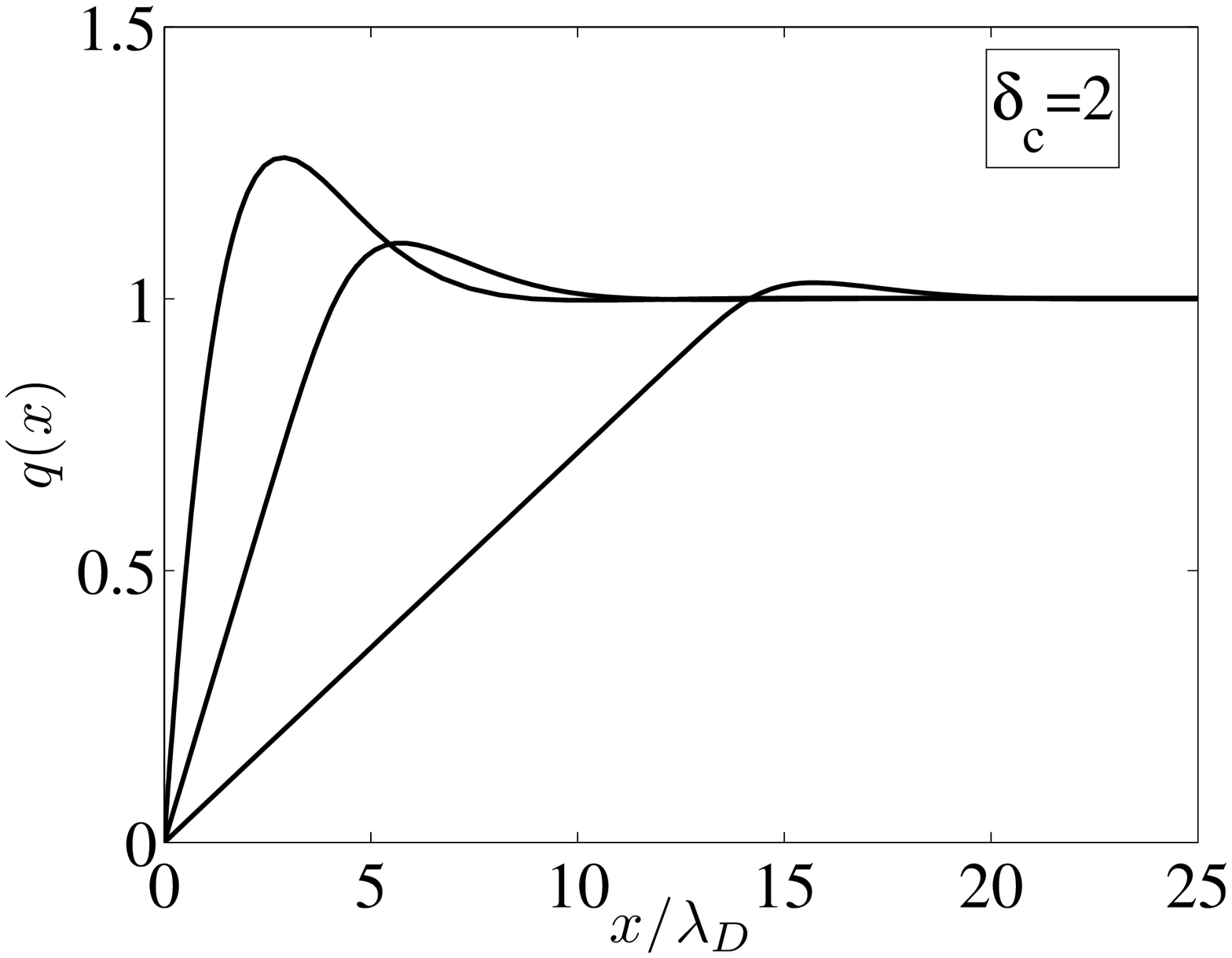}
 \includegraphics[width=2.2in]{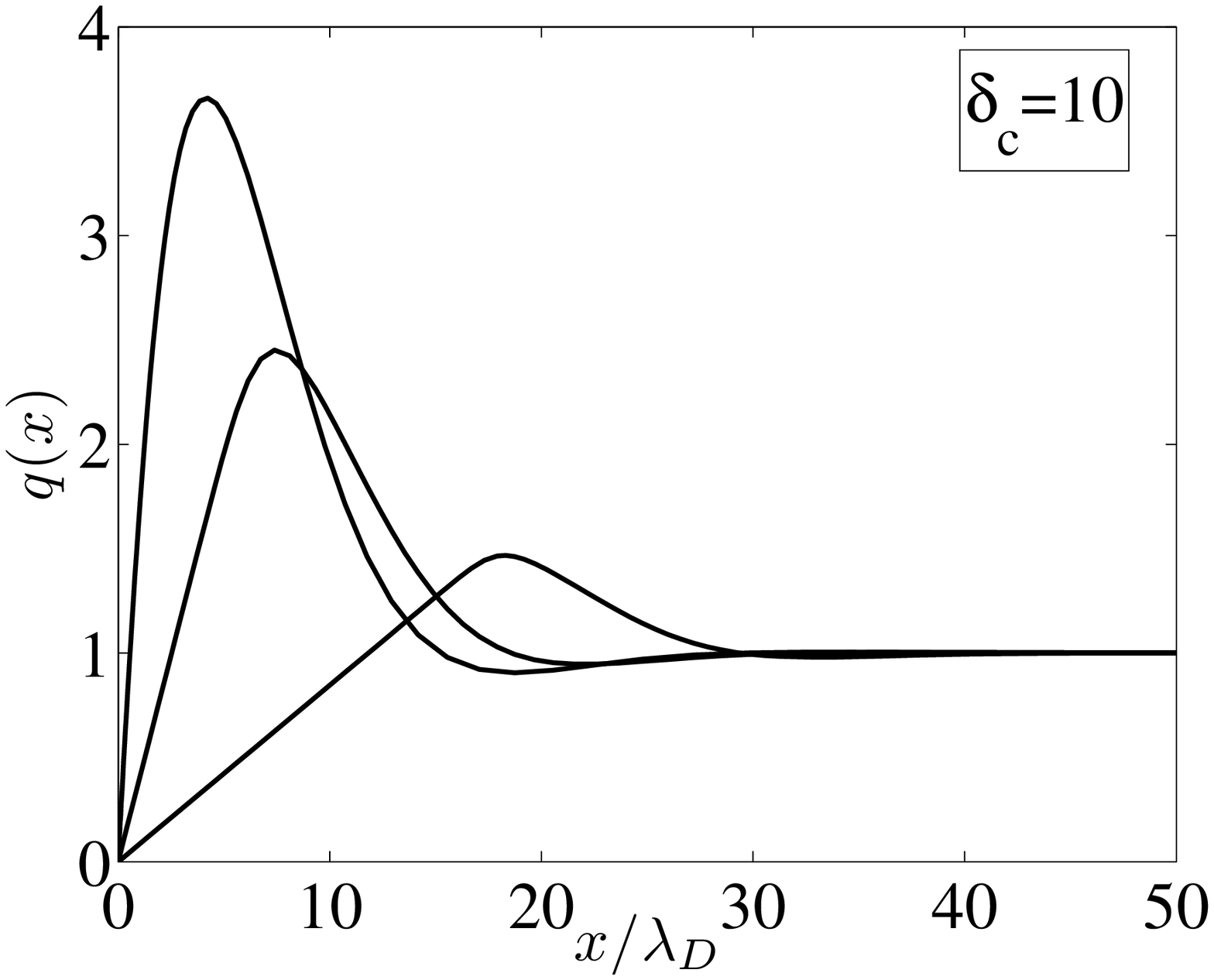}
 \includegraphics[width=2.3in]{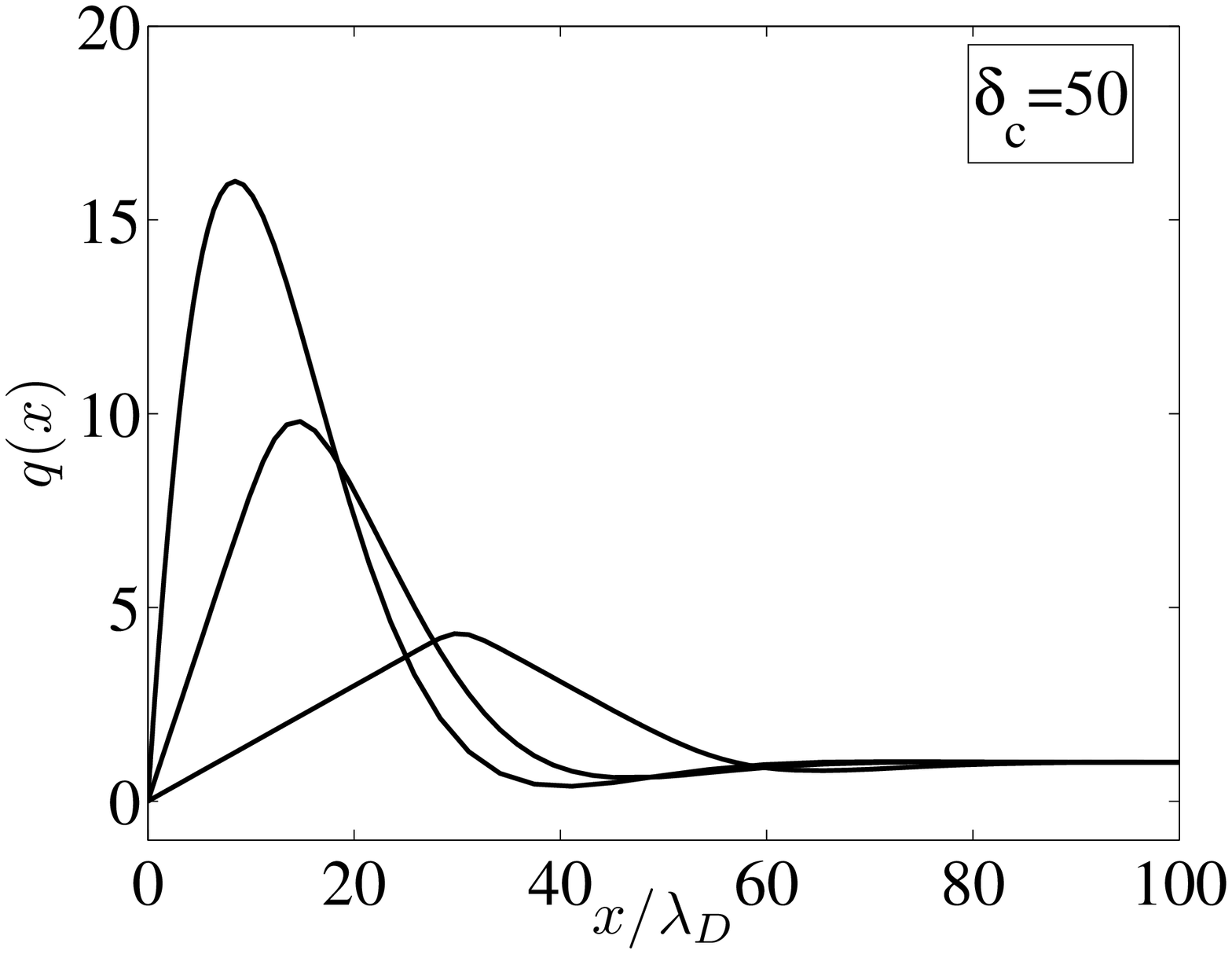}
\caption{Cumulative charge density profiles as a function of distance from the electrode. The charge is integrated cumulatively in
 space and normalized by the total double layer charge; namely
 $q(x) = \int^x_0 \rho(\hat{x}) d \hat{x}/\int^\infty_0 \rho(\hat{x}) d \hat{x}$.
Solutions are shown for $\gamma=1$ and $\delta_c=2,~10~\mathrm{and}~50$, as indicated.
 Applied voltages are $\tilde{V}=1,10$ and $100$ measured in units of $k_B T/e \approx 25$ mV. These cumulative profiles clearly show over-screening. }
\label{fig:rho_integrated}
\end{figure*}

\section{ Low-voltage analytical solution}
The solution to the equations must be calculated numerically.
However, at low voltage we can obtain an analytical solution.
At low voltage, we have  the approximation that,
\begin{equation}
\left( \delta_c^2 \frac{d^4 \tilde{\phi}}{d \tilde{x}^4} - \frac{d^2 \tilde{\phi}}{d \tilde{x}^2}\right) =  -\tilde{\phi}.
\end{equation}
%For the remainder of the Supplementary Material
% we drop the tilde notation used in the main text to denote dimensionless variables
%nd all results hereafter are only stated in dimensionless form.
The analytical solution to this equation depends on whether $\delta_c$ is greater than, equal to, or less that $\frac{1}{2}$. Since $\delta_c$ is presumed large in case of ionic liquids, we present the analytical solution for $\delta_c>\frac{1}{2}$,
\begin{equation}
\tilde{\phi}(x) = \tilde{V}  e^{-k_1 x}  \left( \mathrm{cos}(k_2 x) + A ~\mathrm{sin}( k_2 x) \right),
\end{equation}
where
\[
k_1 = \frac{\sqrt{2 \delta_c + 1}}{2 \delta_c}, ~~ k_2 = \frac{\sqrt{2 \delta_c - 1}}{2 \delta_c}, ~~A  = -\frac{\sqrt{2 \delta_c + 1} (\delta_c -1)}{\sqrt{2 \delta_c - 1} (\delta_c + 1) }.
\]
The total charge in the diffuse double layer can be evaluated from
\[
q = -\int_0^\infty \tilde{\phi} d\tilde{x} = 
\left. \delta_c^2 \frac{\partial^3 \tilde{\phi}}{\partial \tilde{x}^3} \right|_{\tilde{x}=0} -
 \left. \frac{\partial \tilde{\phi}}{\partial \tilde{x}}\right|_{\tilde{x}=0} = -\tilde{V} \frac{\sqrt{2 \delta_c + 1}}{ \delta_c+1}.
\]
The diffuse layer capacitance  in the limit when $\delta_c$ is large is
 approximately  $\tilde{C}_{d}=\sqrt{2/\delta_c}$.
 The diffuse layer capacitance is less than the classical theory without correlations and decreases with the square root of $\delta_c$.

\section{ High-voltage composite approximation }
In ionic liquids, the parameter $\gamma$ is  on the order of unity and
 excluded volume effects are significant.
At voltages beyond the linear response we find that  a condensed layer of counter-ions  forms near the wall.
In this condensed  region close the wall  we could solve,
\[
\left( \delta_c^2 \frac{d^4 \tilde{\phi}}{d \tilde{x}^4} - \frac{d^2 \tilde{\phi}}{d \tilde{x}^2}\right) =  \tilde{\rho}_{\mathrm{max}}
\]
where we  assume that the charge density is a constant, $\tilde{\rho}_{\mathrm{max}}$,
and has reached the maximum value
defined by the value of $\gamma$; i.e. $\tilde{\rho}_{max} = 1/\gamma$ if we apply a negative voltage.
In order to further simplify the approximation, we can assume that
in ionic liquids, $\delta_c$ is typically large and  we solve as an approximation,
\begin{equation}
\delta_c^2 \frac{d^4 \tilde{\phi}}{d \tilde{x}^4} =  \tilde{\rho}_{\mathrm{max}},
\end{equation}
in the wall region.

The general solution for the potential in the wall region becomes a fourth order polynomial.
Using the boundary conditions that  we have fixed voltage $\tilde{V}$ at $x=0$, along with
$\partial^3 \tilde{\phi}/\partial \tilde{x}^3=0$ at the wall, our solution for the potential
has a simple form,
\[
\tilde{\phi}(x) = \frac{\tilde{\rho}_{max}}{24 \delta_c^2} \tilde{x}^4  + B \tilde{x}^2 + C \tilde{x} + \tilde{V}
\]

This polynomial solution which is valid near the wall
can be matched to the low voltage solution provided in the
previous section. Ensuring continuity of the potential, the charge density, and all the
derivatives allows us to solve for the unknown constants of integration. The resulting analysis yields a quartic
equation for the size of the condensed layer.
Once the size of the condensed layer is known, all the constants for the matching
are easy to obtain.
The approximate composite model was found to match the full numerical simulation
as long as the voltage was low enough that a second condensed layer of opposite charge did not begin to form (see
Fig. 1c at $\tilde{V}=100$).

While this analysis may be useful, the resulting quartic equation does
not provide a simple form for the double layer capacitance.
A much simpler form of this composite solution emerges if we make the  additional
approximation that all of the voltage drop occurs across the condensed layer
  and, after the condensed layer, the potential and all its derivatives go to zero.
While  this is not true, it is found from the numerical solutions
to provide a reasonable prediction of the capacitance at "intermediate voltages", where  a condensed layer forms, but not so high that a second condensed layer of opposite charge forms due to over-screening.
While this assumption does not yield complete charge density profiles,
it does provide a useful approximation for the potential in the condensed
layer and thus the capacitance.

\begin{figure}[t]
%a)  \includegraphics[width=2.8in]{fig.cap.fnu.eps}
%b)  \includegraphics[width=2.8in]{fig.cap.fd.eps}\\ \vspace{0.05in}
a)  \includegraphics[width=2.8in]{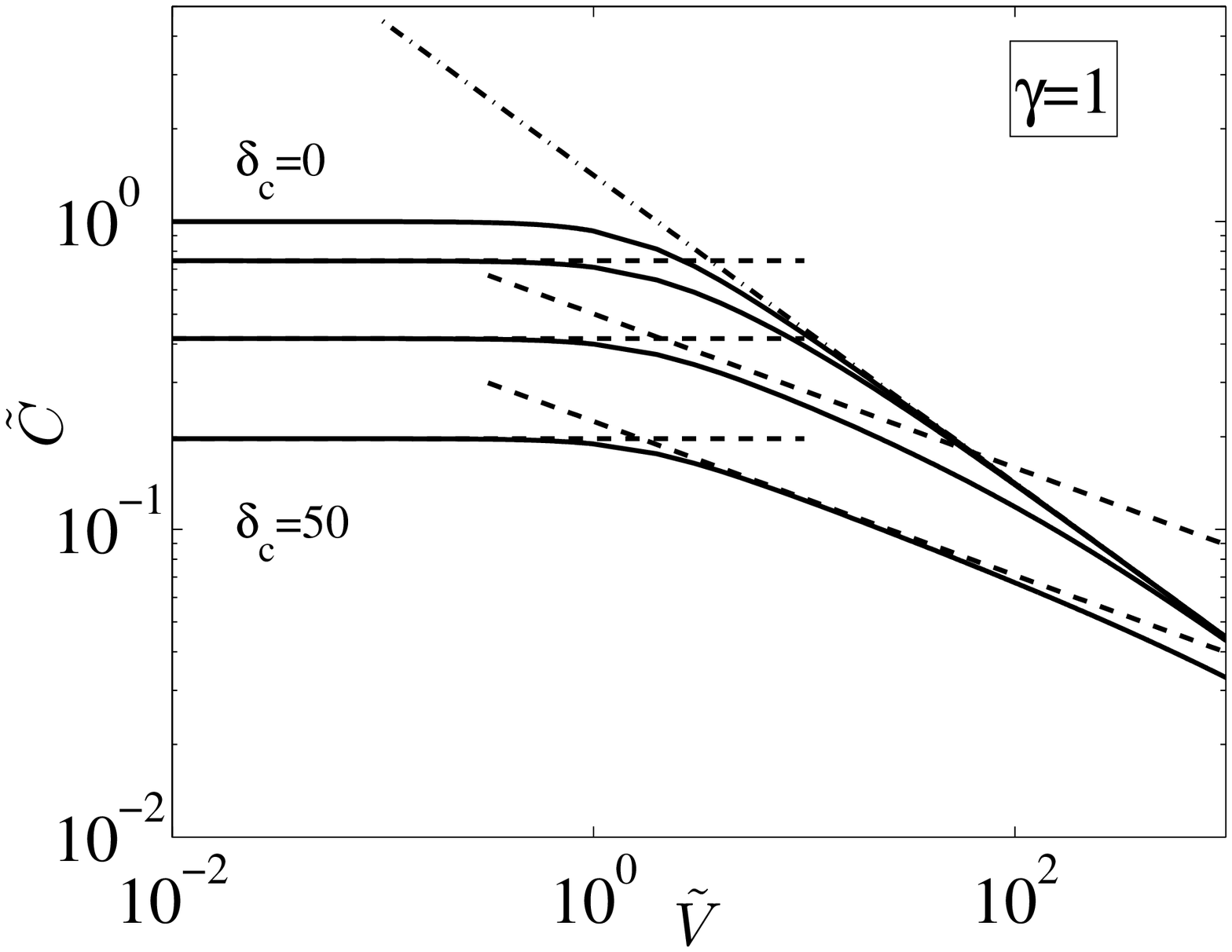}\\
b)   \includegraphics[width=2.8in]{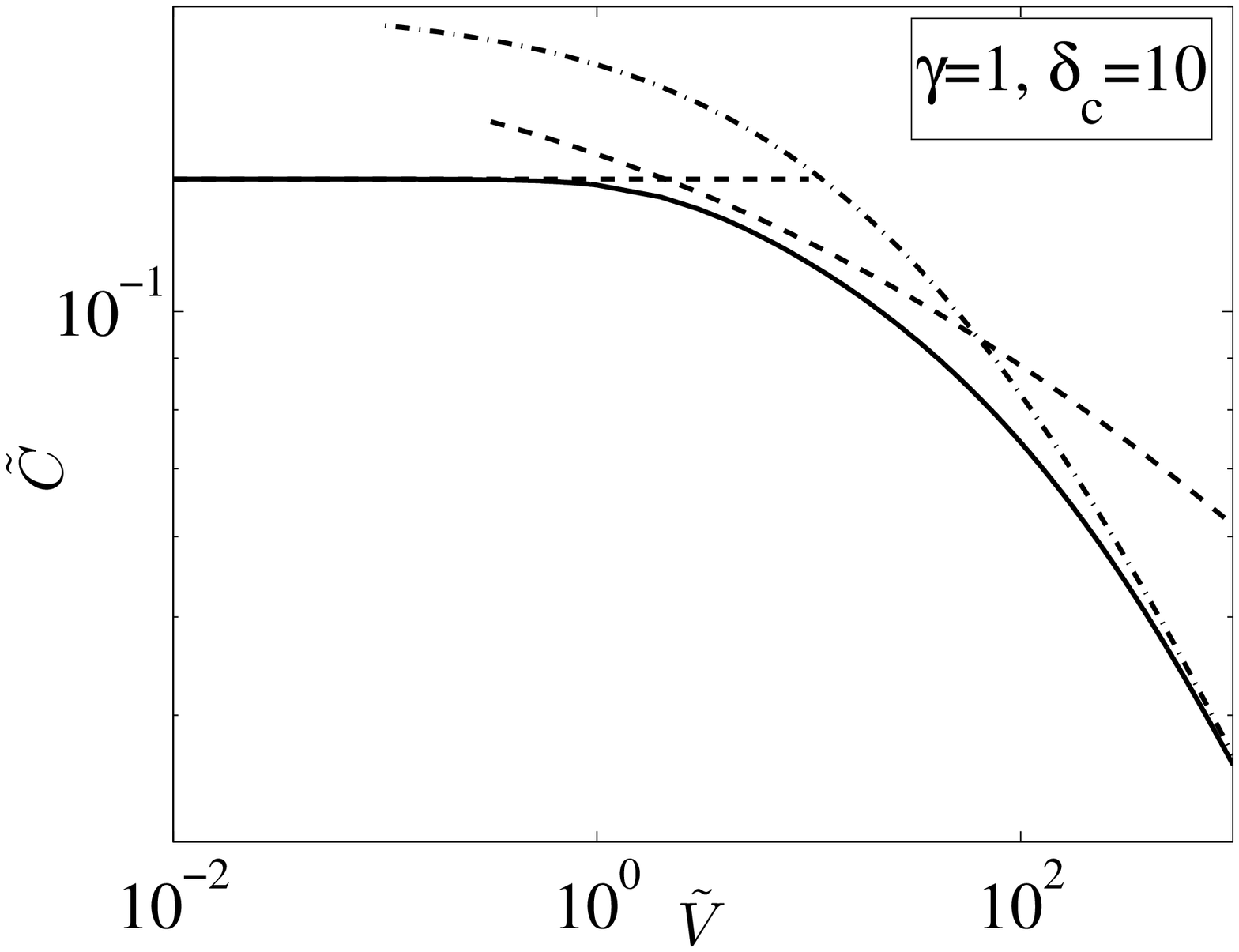}
\caption{(a) Capacitance of the  diffuse double layer,
normalized by the Debye value, as a function of voltage
for $\delta_c=0,~2,~10,$ and $50$ while holding $\gamma=1$.
Numerical solutions (solid curve) are compared with the low-voltage and moderate voltage approximations in Eqs. (7) and (8) of the main text (dashed lines) and  the  high-voltage scaling, $\tilde{C}_{d}\sim \sqrt{2/\gamma \tilde{V}}$ (dash-dot line).
We show only the diffuse layer capacitance to highlight the different scaling laws. 
(b) Total capacitance of the double layer
for $\delta_c=10$ and $\gamma=1$; the
 inner layer is included in series with diffuse layer.
 }
\label{fig:cap}
\end{figure}

Returning to the general solution and using
the simpified boundary conditions,
$\tilde{\phi}(\tilde{x}=L) = 0$ and $\left. \partial \tilde{\phi}/\partial \tilde{x}\right|_{\tilde{x}=L}=0$, we find,
\begin{eqnarray}
\tilde{\phi}(\tilde{x}) = && \frac{-1}{24 \gamma \delta_c^2 } (\tilde{x}^4-L^3 \tilde{x}) +  \nonumber \\
&& \left(\frac{\tilde{V}}{L^2} + \frac{L^2}{8 \gamma \delta_c^2 }\right) (\tilde{x}^2 - L \tilde{x}) + \tilde{V}\left( 1 - \frac{\tilde{x}}{L}\right) \nonumber.
\end{eqnarray}
Setting the second derivative to zero at $\tilde{x}=L$ yields the size of the condensed layer, 
\[
L  = (\tilde{V} \gamma \delta_c^2 8)^{\frac{1}{4}}.
\]
Solving for the total charge $q = \left. \frac{\partial \tilde{\phi}}{\partial \tilde{x}}\right|_{\tilde{x}=0}$, we obtain a simple approximation for the diffuse layer differential capacitance ($\tilde{C}_{d} = d\tilde{Q}/d\tilde{V}$),
\begin{equation}
\tilde{C}_{d}  \sim  \frac{ 8/3 }{   (8 \delta_c^2 \gamma \tilde{V})^{\frac{1}{4}} }
\end{equation}
The scaling presented above is essentially valid at large $\delta_c$ and moderately large voltages.
At high voltage, correlations become irrelevant, since ``crowding beats overscreening", and the capacitance is determined by the excluded volume effects only and has a  scaling  $C_{d} \sim \sqrt{2/\gamma \tilde{V}}$ as previously discovered.

The transitions between these three regimes are evident in Fig.
\ref{fig:cap} where we compare the numerical solution for the capacitance to the simple scaling
laws derived above.  The range of validity of the intermediate voltage expression
simply comes from the intersection of the three regimes.
In Fig. \ref{fig:cap}a we show the diffuse layer capacitance only to clearly show the results of the
simple scaling laws.
At $\delta_c=10$ we find a short transition regime where the $V^{-1/4}$ scaling appears, and for $\delta_c> 50$ the scaling is valid over a wide range.
All the capacitance curves appear to converge to the $\delta_c=0$ solution at high voltages, as expected from our analysis.
When $\delta_c$ is large, however, the voltages where the capacitance curves converge  are quite extreme, so this limiting behavior may have limited applicability. Perhaps it could be used to validate simulations.

In Fig. \ref{fig:cap}b we show the total capacitance (inner layer and diffuse layer in series) for
$\delta_c=10$ and $\gamma=1$. Here we find that the simple scaling laws are useful for understanding the
capacitance computed from the numerical solution of our model.

\end{document}